\documentclass[5p]{elsarticle}
\pdfoutput=1
\usepackage{graphicx}
\usepackage{amsmath} 
\usepackage{amsfonts}
\usepackage{amssymb}
\usepackage{hyperref}
\usepackage{subfigure}
\biboptions{sort&compress}
\usepackage[normalem]{ulem} 

\newcommand{\Trace}{\operatorname{Tr}}

\begin{document}

\title{Filtered topological structure of the QCD vacuum: effects of dynamical quarks}
\author{Falk Bruckmann}
\ead{falk.bruckmann@physik.uni-regensburg.de}
\author{Florian Gruber}
\ead{florian.gruber@physik.uni-regensburg.de}
\author{Andreas Sch\"afer}
\ead{andreas.schaefer@physik.uni-regensburg.de}
\address{Institut f\"ur Theoretische Physik, Universit\"at Regensburg,D-93040 Regensburg, Germany}

\date{\today}

\begin{abstract}
 We systematically compare filtering methods used to extract topological structures on $SU(3)$ lattice configurations. We show that there is a strong correlation of the topological charge densities obtained by APE and Stout smearing. To get rid of artifacts of these methods, we analyse structures that are also seen by Laplace filtering and indeed identify artifacts for strong smearing. The topological charge density in this combined analysis is more fragmented in the presence of dynamical quarks. A power law exponent that characterises the distribution of filtered topological clusters turns out to be not far off the values of an instanton gas model. 
\end{abstract}

\maketitle

\section{Introduction}
The distribution of topological charge density is a characteristic signal for topological objects and their appearance in the QCD vacuum. It is interesting to study also because the topological susceptibility is related to the $\eta'$-mass. 

Such non-perturbative aspects of QCD are best studied through lattice simulations. However, the topological density in those configurations is -- like many other observables -- dominated by quantum fluctuations at the scale of the lattice spacing.
Many methods have been developed to extract the IR content from lattice configurations.
Among them are cooling/smearing, Laplace filtering (for details see below) and the fermionic definition of the topological charge via eigenmodes of a chiral Dirac operator \cite{Niedermayer1999,Horvath:2003yj}. Unfortunately, all these methods introduce their own ambiguities and parameters, which may lead to wrong conclusions.
Cooling for instance is biased towards classical solutions after many steps.

Hence, in order to get a coherent picture of the topological structure of the QCD vacuum, it is necessary to control or even remove these ambiguities. A way to achieve this is the comparison of different filtering methods. In an earlier work \cite{Bruckmann2007c} we have demonstrated that APE smearing, Laplace filtering and Dirac filtering can indeed be matched to each other on quenched $SU(2)$ configurations. Moreover, they agree very well on the `hot spots' in the filtered topological charge, which in addition are subject to an interesting power-law. A similar comparison using overlap (valence) quarks has been performed in Ref.~\cite{Ilgenfritz:2008ia}.
Concerning the total topological charge from different definitions, the agreement has been shown to become almost perfect in simulations with actions that generate smoother configurations a priori (within the Monte Carlo chain) thereby fixing the topological sector\ \cite{Fukaya:2005cw,Bruckmann2009b}.

In the present Letter we apply the methods of Ref.~\cite{Bruckmann2007c} to $SU(3)$ configurations (which can be done with slight modifications) with the aim to study the difference of topological structures in quenched vs. dynamical configurations.
One might expect that light quarks suppress topological objects due to their topological (near) zero modes.
On the other hand, the Adelaide group has found an increased density of smaller nontrivial field configurations after fitting the filtered topological density to instanton profiles \cite{Moran2008a}. This study, however, uses long smearing without comparison to other filtering methods and staggered quarks, whose topological/chiral properties are questionable.

We will show that APE  and Stout smearing yield very similar topological charge densities, i.e.\ agree on a non-perturbative observable, when the parameters are matched as in perturbative studies \cite{Bernard2000d,Capitani2006e}. Nevertheless, to control the ambiguities it is necessary to complement smearing by an independent filtering method, in our case Laplace filtering. This comparison clearly reveals that smearing has to be used with care not to destroy small topological objects.

Our main conclusion concerning the effect of dynamical quarks is that they lead to a more fragmented structure of the QCD vacuum. This is based on the power-law coefficient and the absolute number of topological clusters common to both methods. Therefore, this statement is free of ambiguities and also does not use a particular model of topological objects in the QCD vacuum.

The Letter is organized as follows. In the next two sections we briefly review the definitions of the filtering methods and the topological density we use.  Sect.~\ref{sec: Comparing the methods} deals with the comparison of the filtering methods and how they can be matched optimally. In Sect.~\ref{sec: cluster analysis} we analyze the filtered topological clusters and reveal differences in quenched vs.\ dynamical configurations. Finally, we give our conclusions.

\section{Filtering methods}
One of the first attempts to filter out the UV `noise' has been \emph{APE smearing} \cite{Albanese:1987ds,Falcioni1985}, defined as:
\begin{equation}
\text{\includegraphics[width=0.42\textwidth]{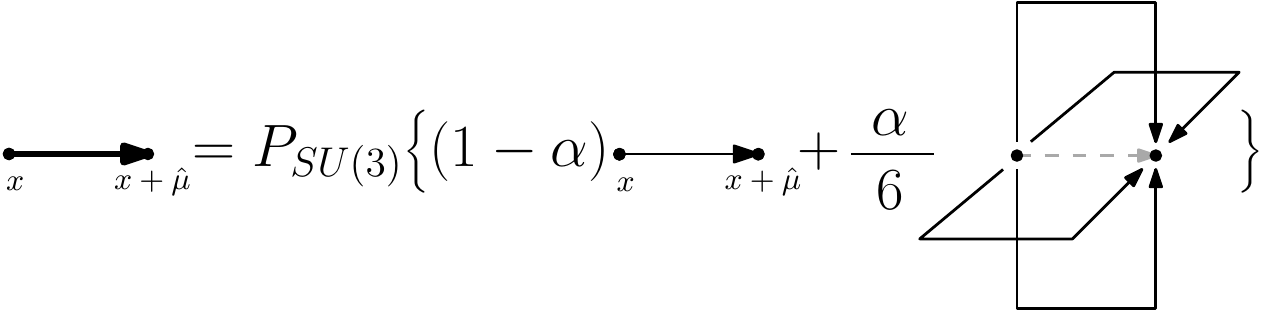}}
\end{equation}
where $\alpha$ determines the weight of the old link and the sum of the attached staples. Throughout this work we have used 4-dimensional APE smearing with the standard value $\alpha=0.45$.

The right-hand side has to be projected back to the gauge group. The projection onto $SU(3)$ is not unique, but a common choice is to define $P_{SU(3)}(W)$ as the that element of the group $V \in SU(3)$ that maximizes $\operatorname{Re}\Trace\{VW^{\dagger}\}$. 
In our work, the maximum is found iteratively using code included in the CHROMA software package for lattice QCD \cite{Edwards2005}.

\emph{Stout smearing} \cite{Morningstar2004a} avoids this projection by using the exponential map:
\begin{equation}
 U_\mu^{ {\rm Stout}}= \exp\Big\{i\; Q_{\mu}(U ,\rho)\Big\}\cdot U_\mu^{{\rm old}},
\end{equation}
where
\begin{equation}
\text{\includegraphics[width=0.43\textwidth]{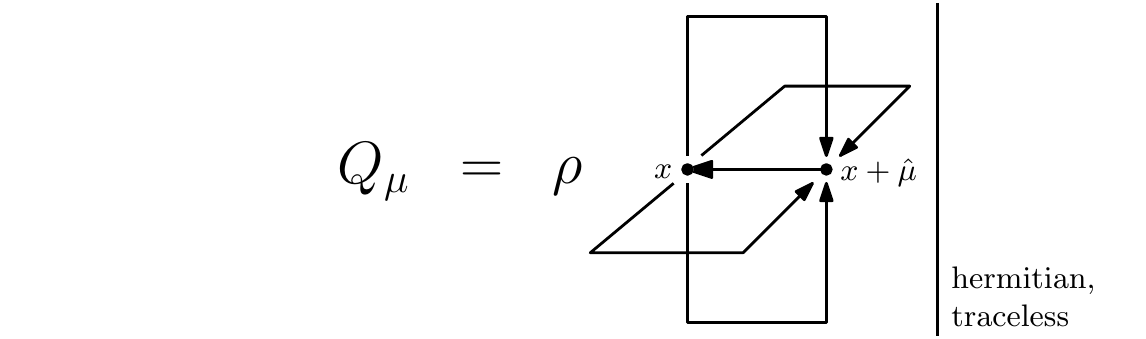}}
\end{equation}
is a hermitian traceless matrix constructed from all plaquettes containing the old link $U_\mu^{\rm old}$ and a smearing parameter $\rho$ \footnote{ Here we also refer to a simultaneous smearing of spatial and temporal links like in the case of APE smearing}. As $e^{iQ} \in SU(3)$ we do not need any projection onto the gauge group. The identification of the hermitian part of a matrix $\Omega$ is not unique,  we use $\tfrac{i}{2}(\Omega^{\dagger}-\Omega)$.
Both smearing methods are iterative procedures, which lead to smoother configurations in each step. These methods had been used with great success in previous studies of the QCD vacuum \cite{DeGrand1998a,Hasenfratz:1998qk,Moran2008a}. But it is hard to decide how many smearing steps one can apply without loosing information about the underlying structure. 

A relatively new and substantially different method is \emph{Laplace
filtering} \cite{Bruckmann2005a}. This method is based on a spectral
decomposition of the links in terms of eigenmodes $\Phi_n(x)$ of the covariant
lattice Laplacian:
\begin{equation}
 U_\mu(x) = -\sum_{n=1}^{N_{tot}}\lambda_{n}\Phi_n(x)\otimes\Phi_n^{\dagger}(x+\hat\mu),
\end{equation} 
where the sum runs over all $N_{tot}\equiv {\rm Vol}\cdot N_c $ eigenmodes. As the eigenvalues are related to some energy scale (squared), one can truncate the sum at $N~\ll~N_{tot}$ to get rid of the high energy contributions:
\begin{equation}
 U_\mu^{\text{Lapl}}(x) = P_{SU(3)}\left\{-\sum_{n=1}^{N}\lambda_{n}\Phi_n(x)\otimes\Phi_n^{\dagger}(x+\hat\mu)\right\}.
\end{equation}
This procedure acts as a low-pass filter in the sense of a Fourier decomposition: with lower values $N$
of included modes the Laplace filtering gets stronger. Taking the filtered links as a starting point, one can in principal measure any observable.

While both strategies aim at smoothing the gauge fields, it should be stressed that Laplace filtering is completely different from smearing, because it is based on extended objects, namely the eigenmodes, and does not locally modify the gauge links in contrast to smearing. 

In this sense this method is quite similar to Dirac filtering
\cite{Horvath:2003yj}, which provides direct access to the topological
charge density via the low-lying eigenmodes of the Dirac operator. However,
Laplacian modes are chirality blind and do not obey an index theorem, which
connects the zero-modes of the (chiral) Dirac operator to the topological
charge.
From a computational point of view Laplace filtering is much cheaper than Dirac filtering (with good chirality properties).

\section{Topological Charge Density}
Throughout this Letter we will focus on the topological charge density:
\begin{equation}
q(x)=\frac{1}{32\pi^2} \epsilon_{\mu\nu\rho\sigma}{\rm Tr}\Big(F_{\mu\nu}(x)F_{\rho\sigma}(x)\Big),
\end{equation}
where $F_{\mu\nu}$ is an improved discretization of the field strength tensor \cite{Bilson-Thompson2003}, which combines $1\times1$, $2\times2$ and $3\times3$ loops to achieve $O(a^4)$-improvement at tree-level\footnote{for details on the improvement coefficients see \cite{Bilson-Thompson2003}}. This definition is better behaved in the continuum limit than the naive discretization with only $1\times1$ loops and gives topological charges $Q=\sum_x q(x)$ closer to integers.

In order to get results for dynamical and quenched configurations which can be easily compared, we have chosen two $SU(3)$ ensembles with the same lattice spacing and the same physical volume (see Tab.~\ref{tab: configurations}). The ensembles were generated with the L\"uscher-Weisz gauge action and a chirally improved Dirac operator \cite{Gattringer:2000qu}. For the dynamical simulations two flavors of mass degenerate light quarks were used (details can be found in \cite{Gattringer2008}). 

\begin{table}[!t]
\centering
 \begin{tabular}{l @{\extracolsep{5mm}} c @{\extracolsep{5mm}}c @{\extracolsep{5mm}}c @{\extracolsep{5mm}}c }
\hline\hline
 & lat. size & lat. spacing & $\beta_{L W}$ \\ \hline
quenched & $16^3\cdot32$ &  0.148 & 7.90 \\ 
dynamical & $16^3\cdot32$ & 0.150 & 4.65 \\ \hline\hline 
\end{tabular}
\caption{Details on the gauge configurations used in this work.}\label{tab: configurations}
\end{table}

\section{Comparison of the Methods}\label{sec: Comparing the methods}
\begin{figure}
\centering
 \subfigure[APE smearing]{\label{fig: Ape slice}
 \includegraphics[trim=0 40 0 50,width= 0.4\textwidth]{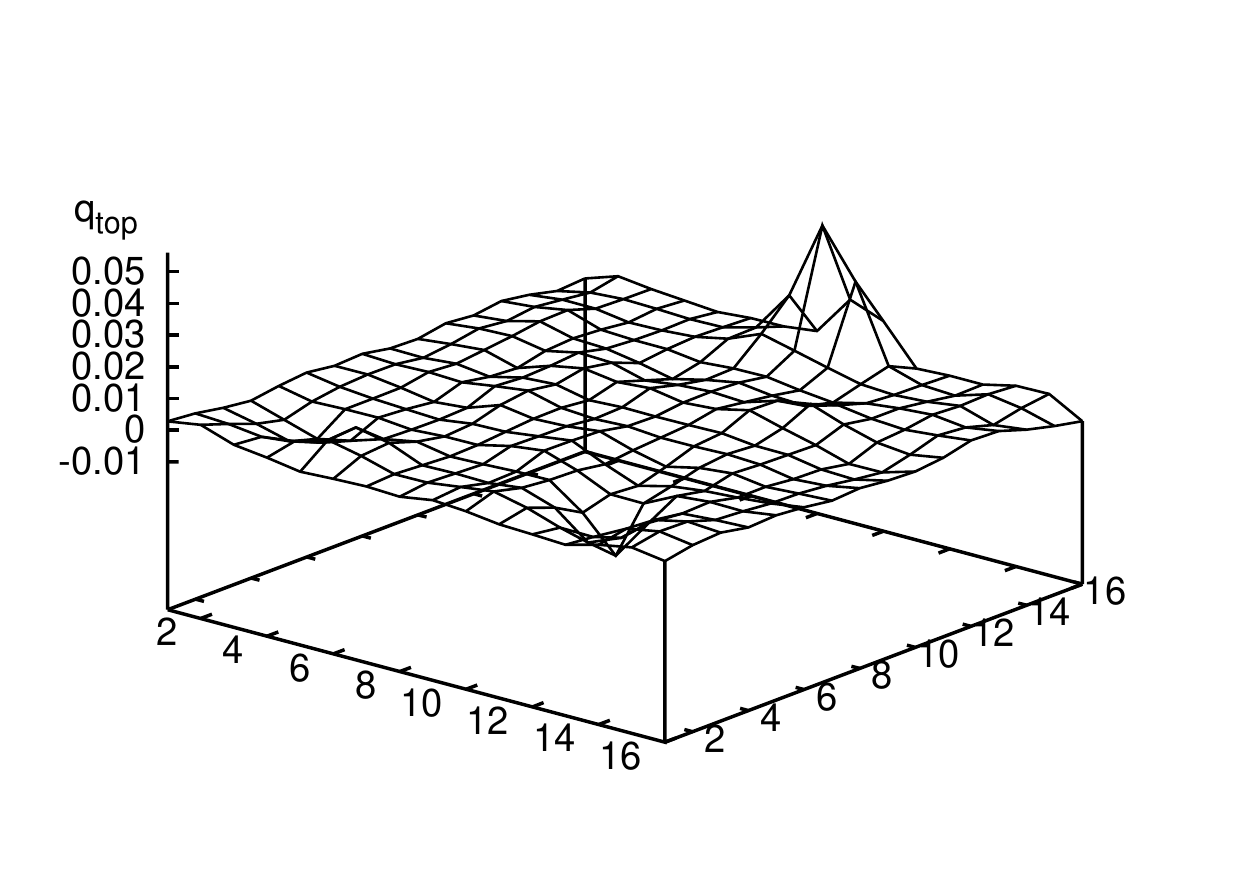}}
\subfigure[Laplace filtering]{\label{fig: Laplace slice}
 \includegraphics[trim=0 40 0 50,width=0.4\textwidth]{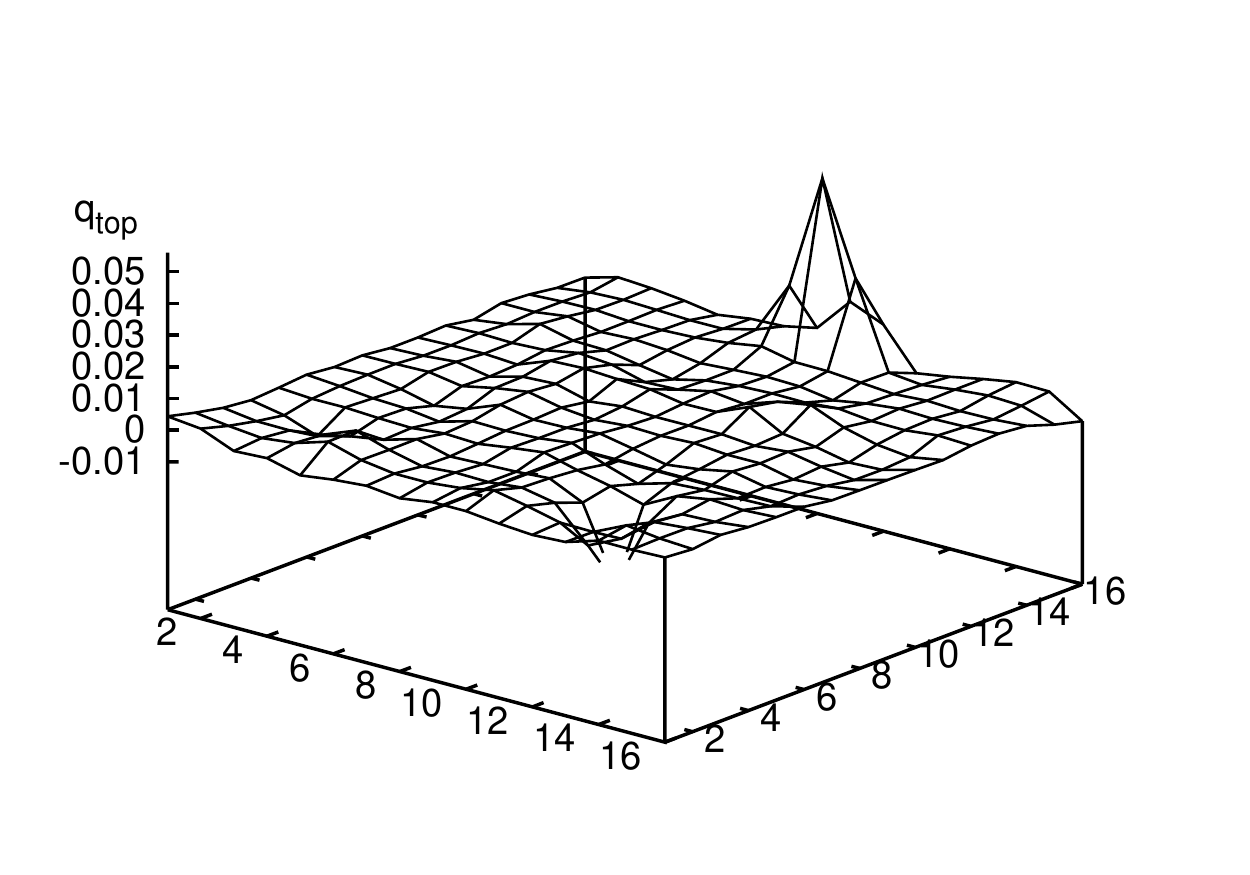}}
\caption{Comparison of the same slice of the topological charge density resulting from 7 APE smearing steps (a) and 500 Laplace modes (b).\label{fig: compare slice}}\vspace*{-12pt}
\end{figure}

For quenched $SU(2)$ lattice configurations we had observed a similarity of the topological density obtained with the different filtering methods \cite{Bruckmann2007c}. A corresponding example of a quenched $SU(3)$ configuration after APE smearing and Laplace filtering is shown in Fig.~\ref{fig: compare slice}.

For a quantitative comparison one needs a measure of the similarity. In \cite{Bruckmann2007c} the following quantity was introduced:
\begin{equation}\label{eq: Xi}
\Xi_{A B}\equiv \frac{ \chi_{A B}^2}{ \chi_{A A}\;\chi_{B B}}
\end{equation}
with
\begin{equation}
\chi_{A B}\equiv\big(1/V\big)\sum_x\;\big(q_A(x)-\overline{q}_A\big)\;\big(q_B(x
)-\overline{q}_B\big)
\end{equation}
being the correlator of two topological charge densities $q_A(x)$ and $q_B(x)$ (mean values $\overline{q}=Q/V$ are subtracted for convenience).

$\Xi$ is a positive quantity and equals $1$, if the densities differ only by a constant scaling factor. The great advantage of this quantity is that we do not have to know the normalization factors of the topological charge densities, since these factors drop out in Eq.~(\ref{eq: Xi}) and we are able to compare only the relative difference at each lattice point.

For a quantitative analysis of the local structure resulting from different methods $A$ and $B$, it is useful to define `best matching' pairs of filter parameters, by maximizing $\Xi_{A B}$.

To be precise, we keep the smearing parameters constant and compute $\Xi_{A B}$ for up to 50 APE steps and 50 Stout steps respectively up to 500 Laplace modes. More smearing steps have not been applied, as too much smearing will destroy the local topological structure, whereas more Laplace modes would be too expensive.

\begin{figure*}[t!!]
\centering
\begin{center}
\begin{tabular}{cc}
\includegraphics[trim=10 0 10 0,width= 0.35\textwidth]{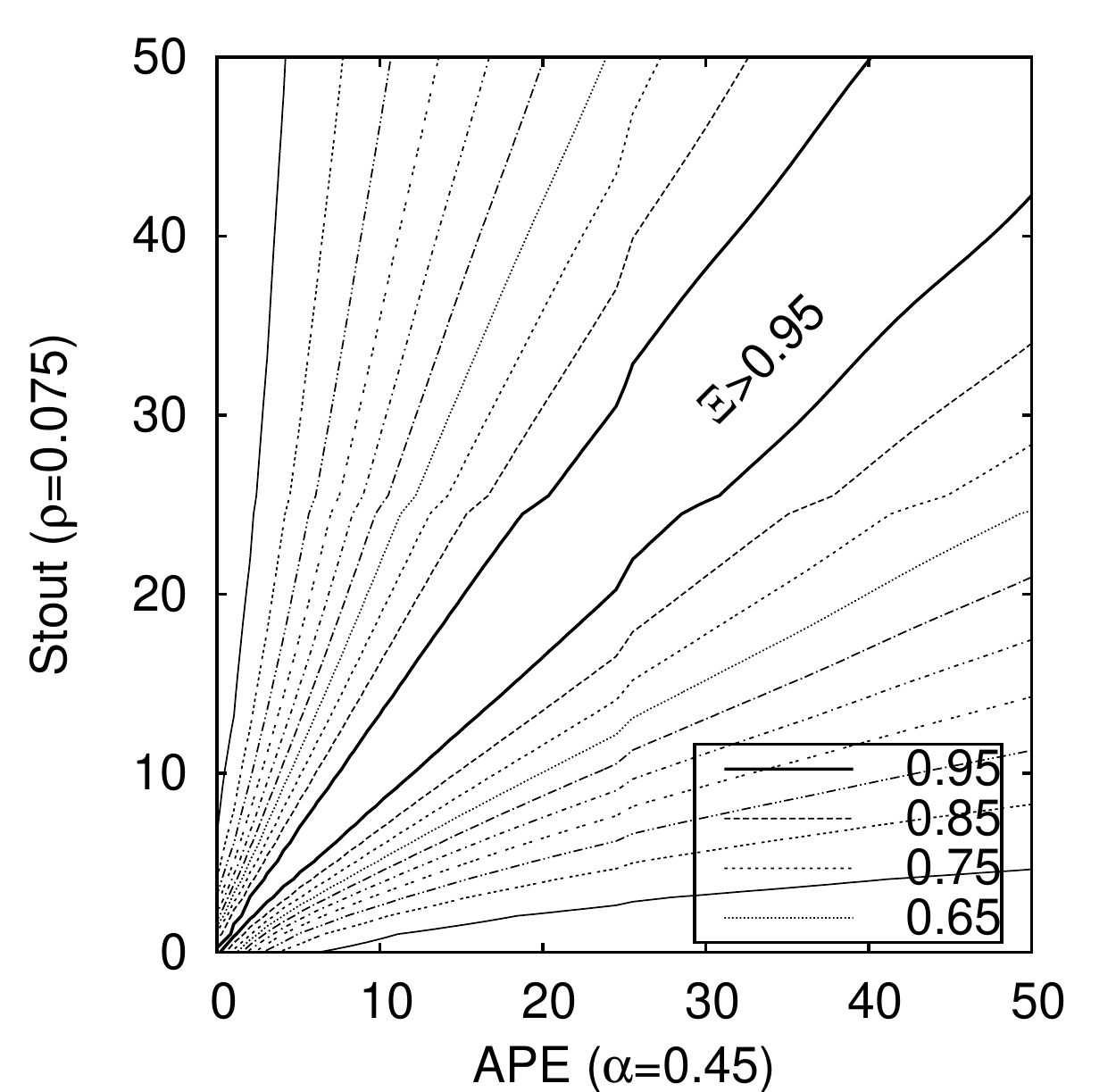} & \includegraphics[trim=10 0 10 0,width=0.35\textwidth]{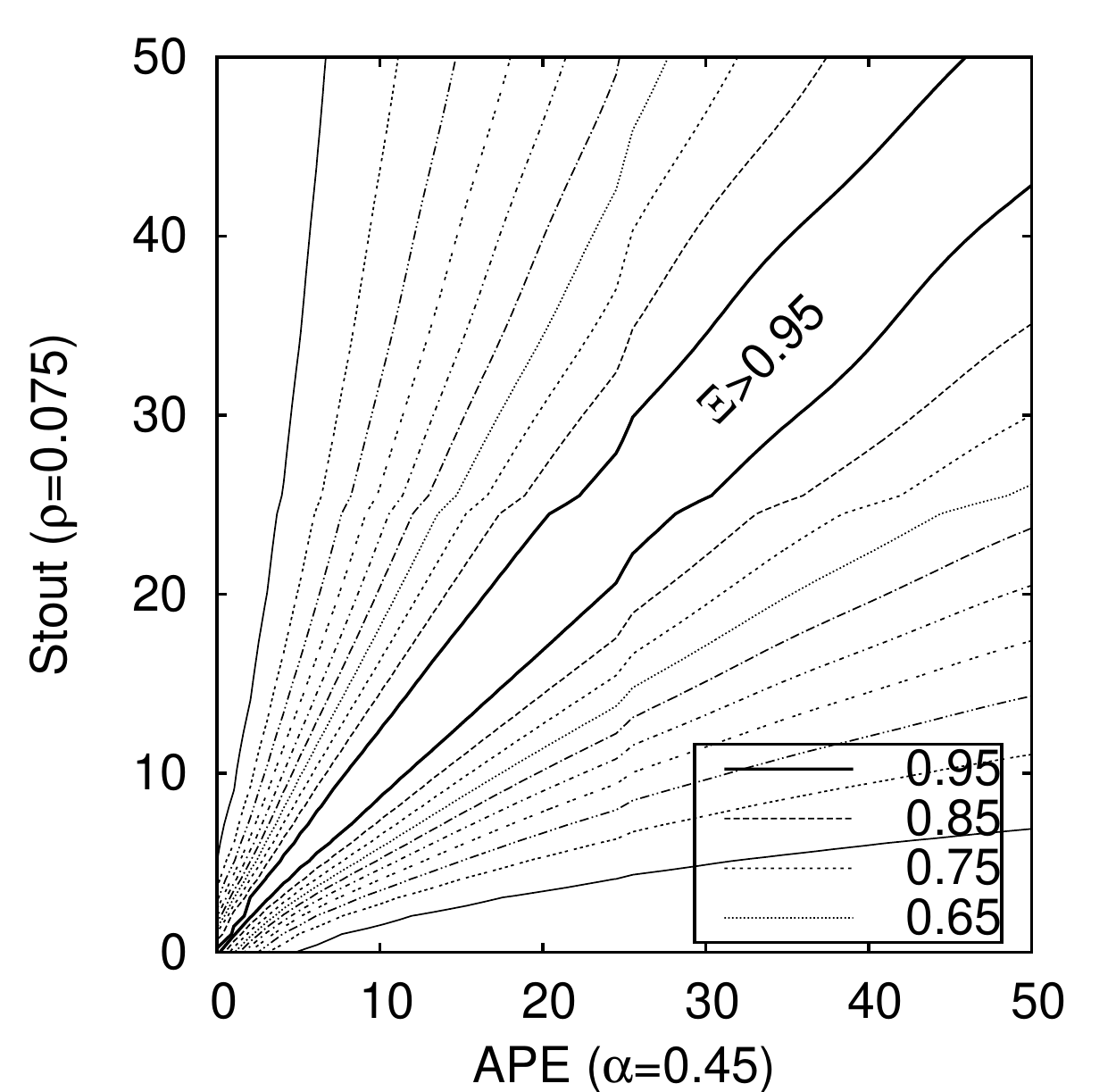} \\ 
quenched & dynamical\\ 
 \includegraphics[trim=10 0 10 0,width=0.35\textwidth]{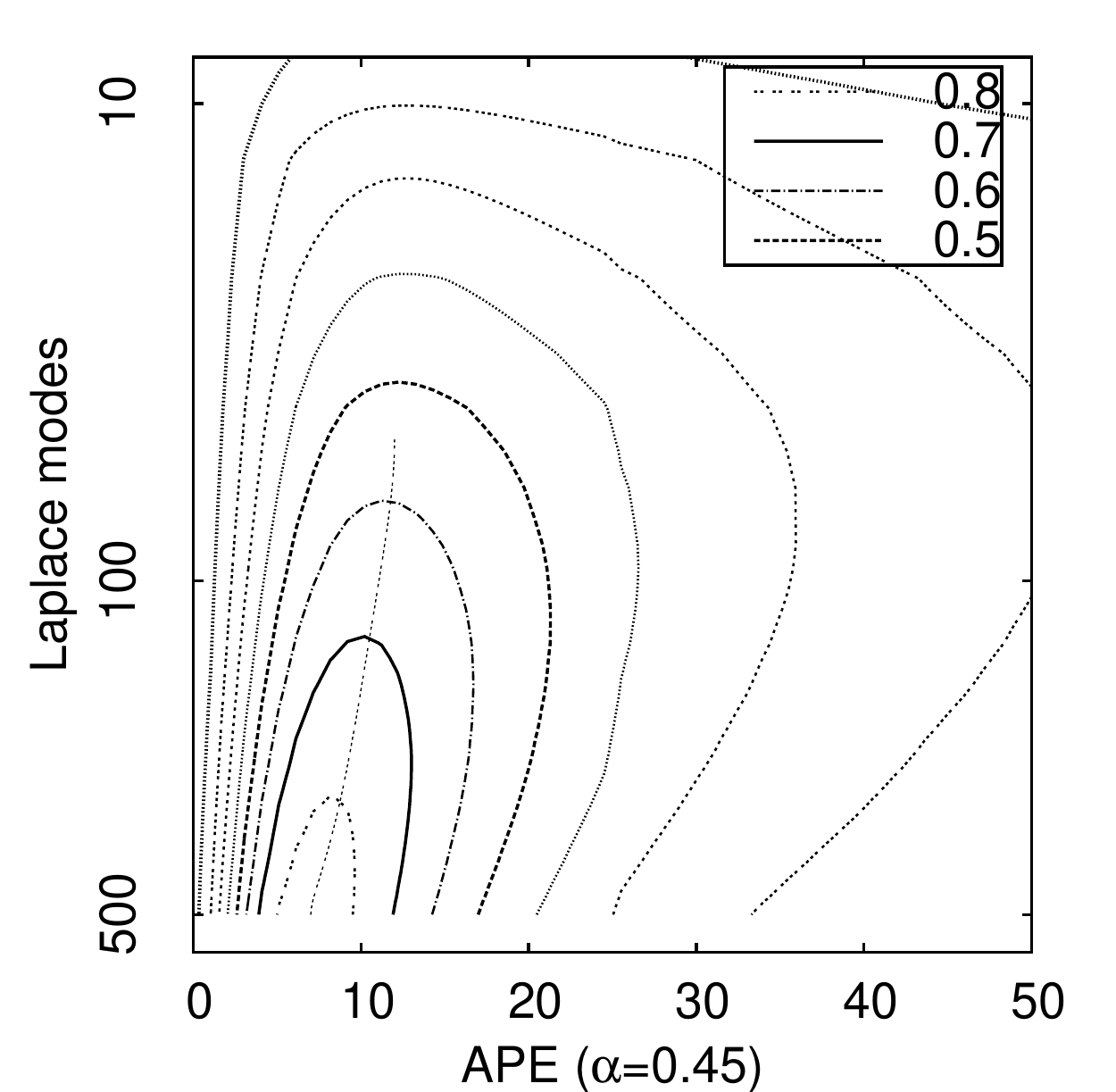}&   \includegraphics[trim=10 0 10 0,width=0.35\textwidth]{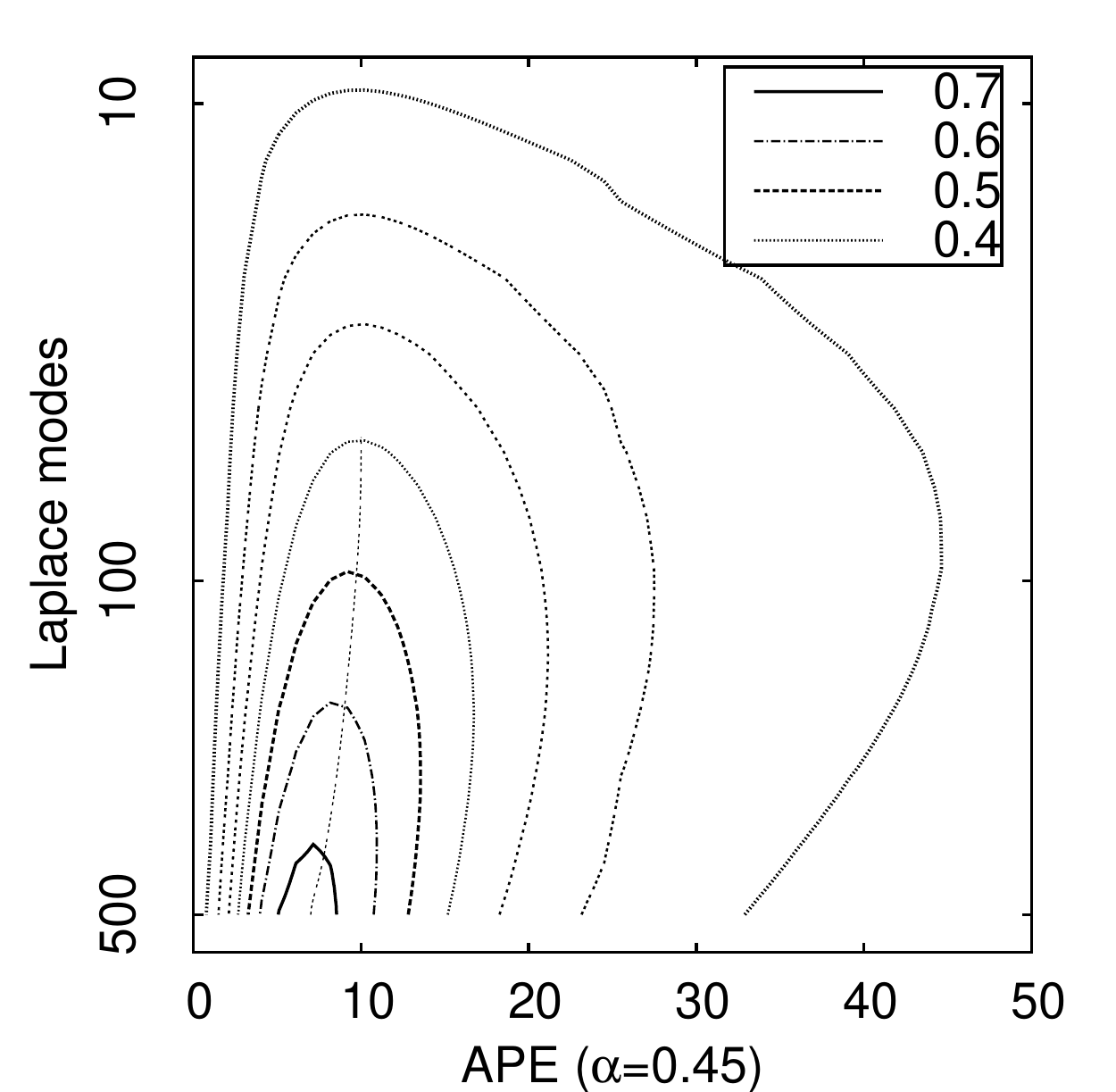}\\ 
 quenched & dynamical 
\end{tabular}
\end{center}
\caption{Comparison of APE  with Stout smearing and APE smearing with Laplace filtering for quenched and dynamical configurations. The degree of similarity $\Xi$ (see the definition (\protect\ref{eq: Xi})) is given by contour lines. Shown are mean values of 80 configurations for APE  vs. Stout smearing and 30 configurations for APE smearing vs. Laplace filtering. The comparisons APE  vs. Stout smearing  for quenched (upper left) and dynamical configurations (upper right) give very similar results, only the band with $\Xi > 0.95$ is slightly broader in the quenched case. The values of $\Xi$ for APE smearing vs. Laplace filtering does not reach such high values and the agreement is worse in the dynamical case (lower row). For mild filtering we have plotted a ridge line of best matching parameters that will be used in the cluster analysis later.\label{fig: Xi}}
\end{figure*}
We first discuss the matching of APE  and Stout smearing, see Fig.~\ref{fig: Xi} upper row. In the absence of smearing (lower left corner) one has  $\Xi=1$ as the configurations are identical.

An interesting observation is that the maximal values of $\Xi$ lie on some kind of ridge line, where $\Xi > 0.95$ within our range of smearing steps. Thus, the different smearing techniques reveal almost the same structures. The slope of this line is determined by the ratio of the smearing parameter.

An almost one-to-one correspondence for the topological density after the same number of APE  and Stout smearing steps (i.e.\ a slope of 1) is achieved for the parameters related as $\alpha=6\cdot\rho$ \cite{Bruckmann2009}.

 This is consistent with the perturbative result of Ref.~\cite{Bernard2000d} and with the non-perturbative result from Capitani {\em et al.}\ \cite{Capitani2006e}. The latter group has focused on rather global observables, which are not related to topology, with up to three smearing steps. Our non-perturbative result reflects the \emph{local} similarity of both methods and their strongly correlated topological charge densities up to 50 steps.

The described equivalence applies to both dynamical and quenched configurations.

The comparison is not so perfect when we consider APE smearing and Laplace filtering, see Fig.~\ref{fig: Xi} lower row. Overall, these plots are similar to Fig.~3 (bottom panel) of Ref.~\cite{Bruckmann2007c}. Although there is a qualitative similarity for quenched and dynamical configurations, we find a quantitative difference. The agreement of both methods is worse in the dynamical case which means that there might be more artifacts from a single method.

Also here we find some kind of ridge line of the best matching values of $\Xi$. Although this ridge is quite pronounced for weak filtering, we are confronted with the problem to identify the best matching parameters in the strong filtering regime, i.e. for less than 50 Laplace modes and more than 20 smearing steps. Furthermore, there is no set of best matching parameters with $\Xi>0.5$ in this regime and thus one can hardly say that both methods agree at all.

As another measure for the similarity of topological landscapes at best matching we use the relative point overlap (RPO) \cite{Bruckmann2007c}:
\begin{equation}\label{eq: RPO}
s_{AB}=
    \displaystyle\sum_{
      \stackrel{x\in X_A \cap X_B}
      {q_A(x)q_B(x)>0}}
    1 \:\:/
    \displaystyle\sum_{x\in X_A\cup X_B}\!\!\!\! 1\,.
\end{equation}
Here $X_A$ is the set of points $x$ with topological density $|q_A(x)|$ above a cut-off which is adjusted such that the volume fraction is the same for both methods $A$ and $B$, $vol(X_A)/Vol=vol(X_B)/Vol=f$. Again, a value $s_{AB}$ close to 1 signals a good agreement between $q_A$ and $q_B$.

\begin{figure}
\centering
\includegraphics[trim= 0 0 0 0,width=0.47\textwidth]{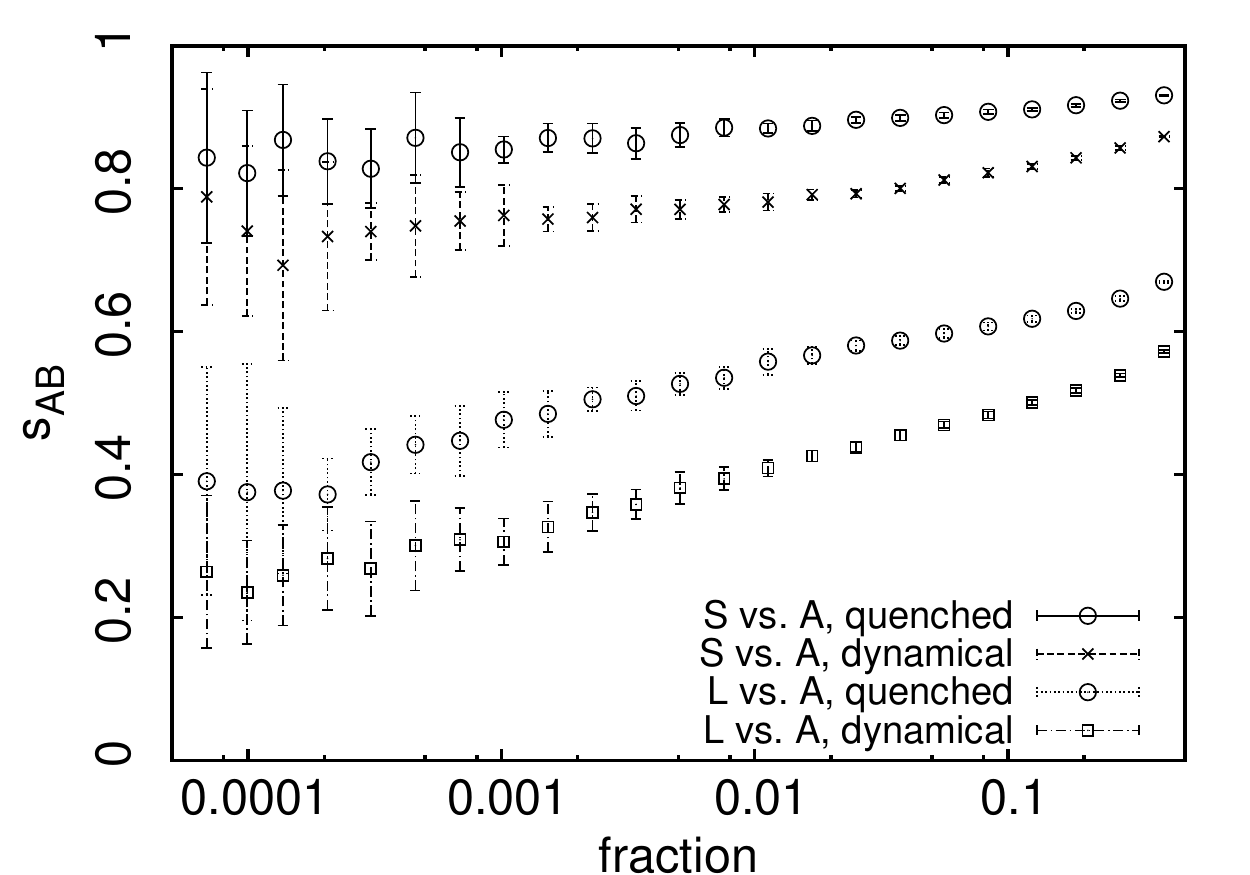}
\caption{Releative point overlap $s_{AB}$ (see Eq.~\ref{eq: RPO}) of APE and Stout smearing (7 iterations each) as well as for APE smearing and Laplace filtering (7 iterations vs. 500 Laplace modes) averaged over 10 configurations each.}\label{fig: RPO}
\end{figure}
The values of the RPOs are shown in Fig.~\ref{fig: RPO}. They are rather constant over a large range of volume fractions $f$. 

The RPO for APE vs.\ Stout smearing is quite large supporting the similarity of the two smearing methods. The RPO for APE smearing and Laplace filtering is smaller and comparable to its values found for quenched SU(2) configurations \cite{Bruckmann2007c}. Again we find the tendency of less agreement in the case of dynamical configurations.

\section{Cluster Analysis of the Topological Charge Density}\label{sec: cluster analysis}

In order to obtain information about the topological structure of the QCD vacuum that can be compared to continuum models, we analyze the cluster structure of the topological charge density. Two lattice points belong to the same cluster, if they are nearest neighbors and have the same sign of the topological charge density.

\begin{figure}
\centering
\includegraphics[trim= 0 0 0 0,width=0.47\textwidth]{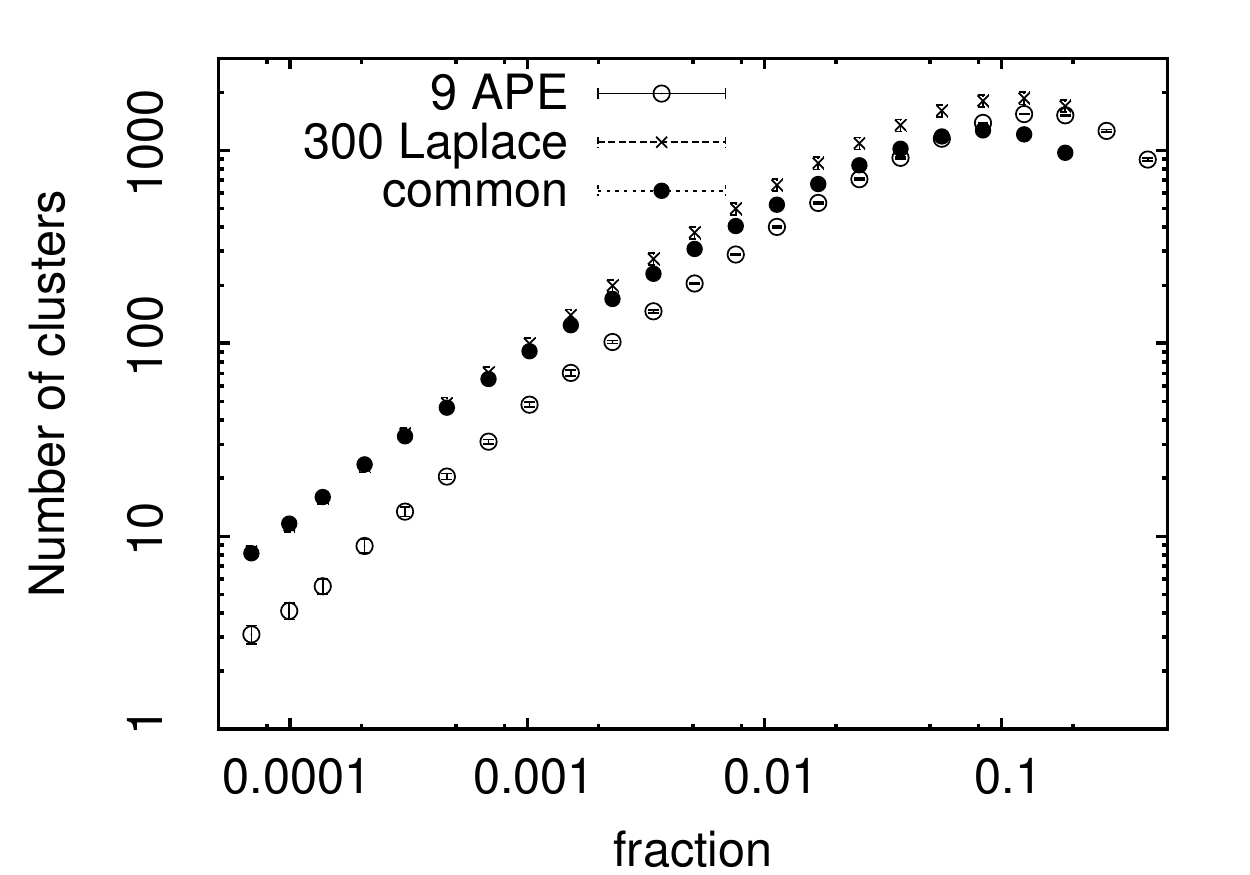}
\caption{Number of clusters as a function of the fraction of all points above a variable cut-off averaged over 30 dynamical configurations. The exponent of the underlying power-law is highly characteristic for the topological structure and directly related to the size distribution of topological objects.}\label{fig: power-law}
\end{figure}

 For such clusters in $SU(2)$ we had found an interesting power law \cite{Bruckmann2007c}. To that end one cuts the absolute value of the topological charge at a (variable) cut-off $q_{\rm cut}$ and considers the number of clusters above this cut-off as a function of the fraction $f$ of all points obeying $|q(x)|>q_{\rm cut}$ w.r.t.\ the total number of lattice points. An example of this power-law for SU(3) can be found in Fig. \ref{fig: power-law}.

The exponent $\xi\equiv {\rm d}\log N_{\rm clust.}(q_{\rm cut})/{\rm d}\log N_{\rm points}(q_{\rm cut})$ of this power law is highly characteristic for the topological structure of the QCD vacuum. Different models lead to different predictions. This allows for a sensitive test.

Pure noise, for instance, yields an exponent $\xi=1$, because every point forms its own cluster. On the other hand, the exponent is close to zero for very smooth densities with large structures, because in such a situation a small change in the cut-off will add new points but not reveal new clusters.

The exponent can be related to the size distribution $d(\rho)\sim \rho^\beta$ of equally charged topological objects with arbitrary shape function and arbitrary dimensionality \cite{Bruckmann2007c}. For instance a dilute gas of instantons would have $\xi=0.64$ in the quenched case and $\xi=0.66$ in the dynamical case ($N_f=2$). 

A cluster analysis has one great advantage. It allows to reduce  ambiguities coming from a single filter by taking only those clusters into account, which are common to different filters. If there is an artifact coming from one method, it is unlikely that this artifact will also be seen by the other one, such that the common structures are almost free of ambiguities.
\begin{figure}
\centering
\subfigure[Smearing]{\label{fig: APE cluster analysis}
{\includegraphics[width=0.437\textwidth]{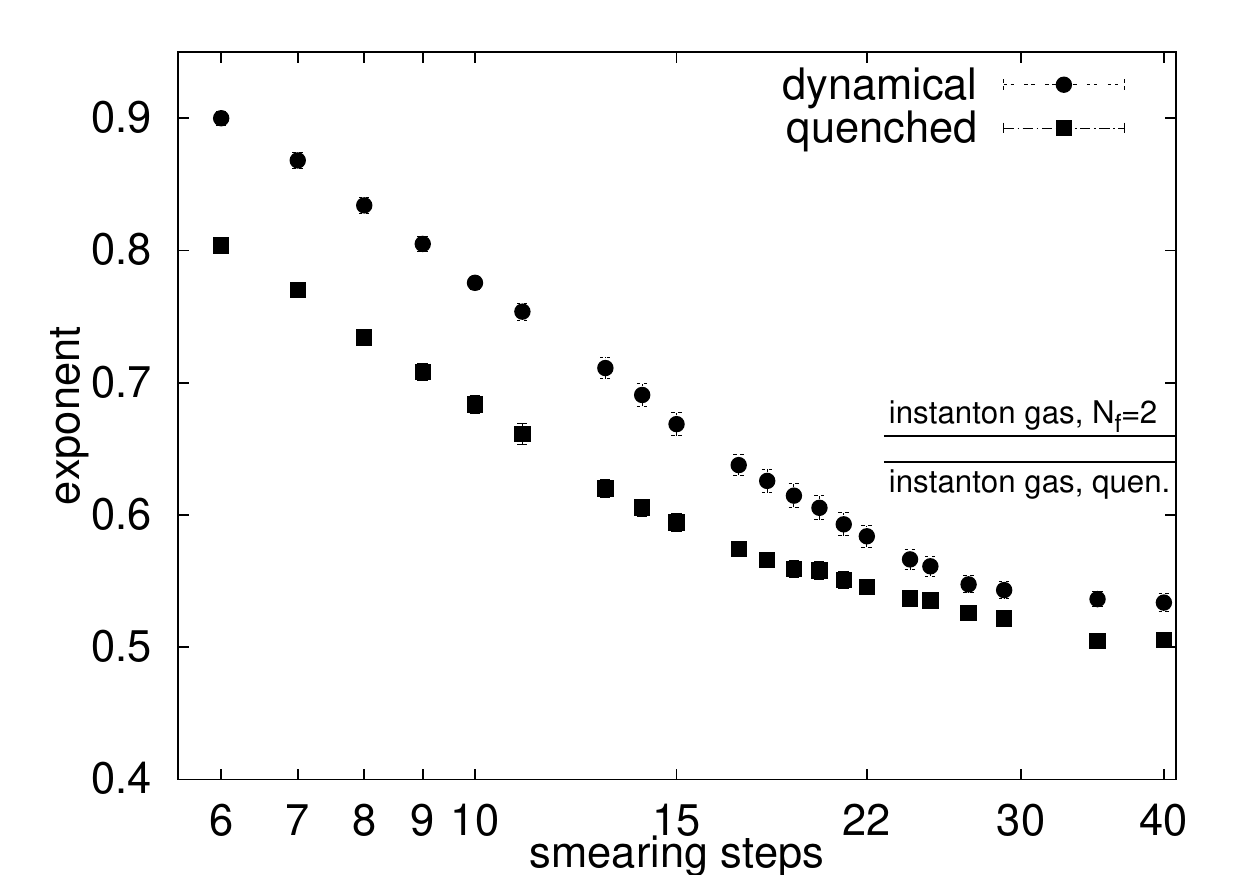}}}
\subfigure[Laplace filtering]{\label{fig: Laplace cluster analysis}
{\includegraphics[width=0.437\textwidth]{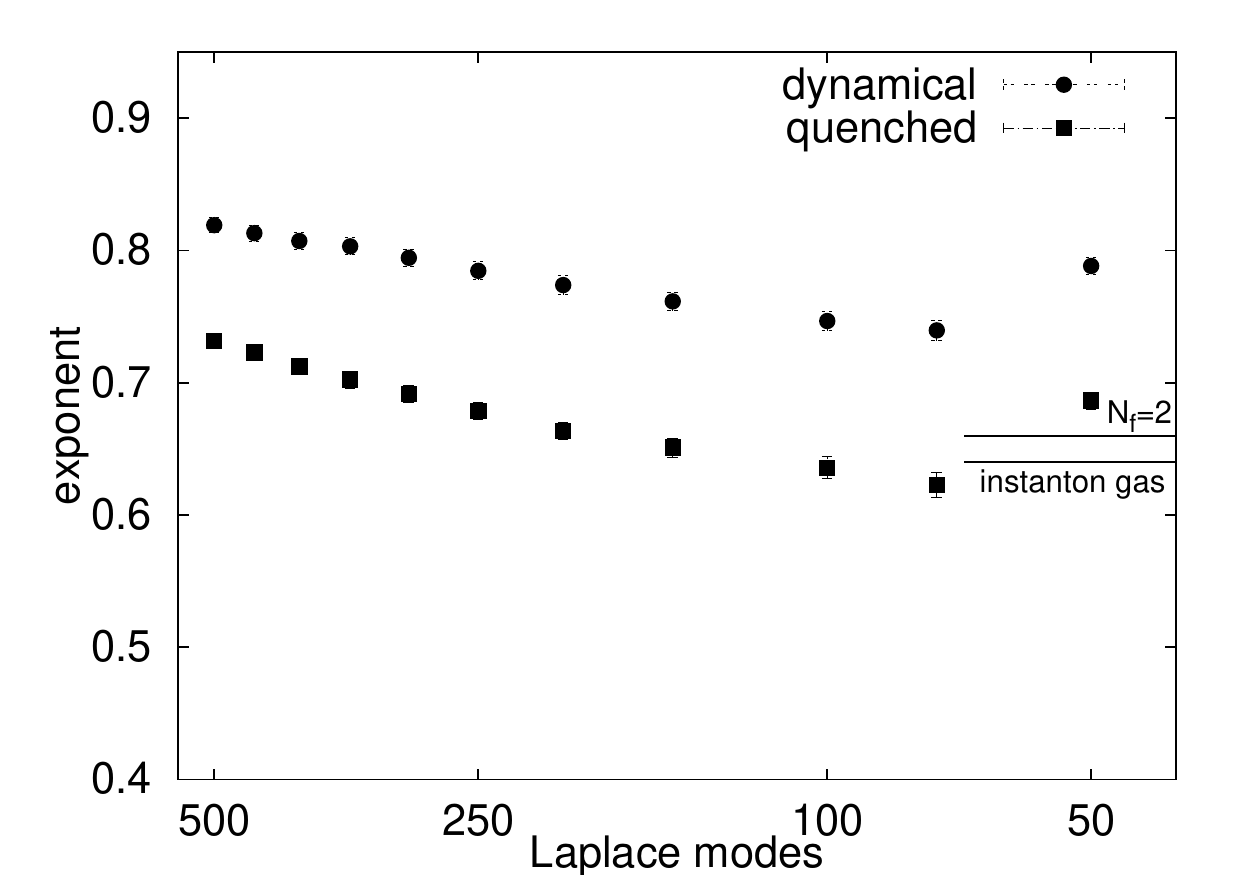}}}
\subfigure[matched Filtering]{\label{fig:Laplace Ape matched cluster analysis}
{\includegraphics[width=0.445\textwidth]{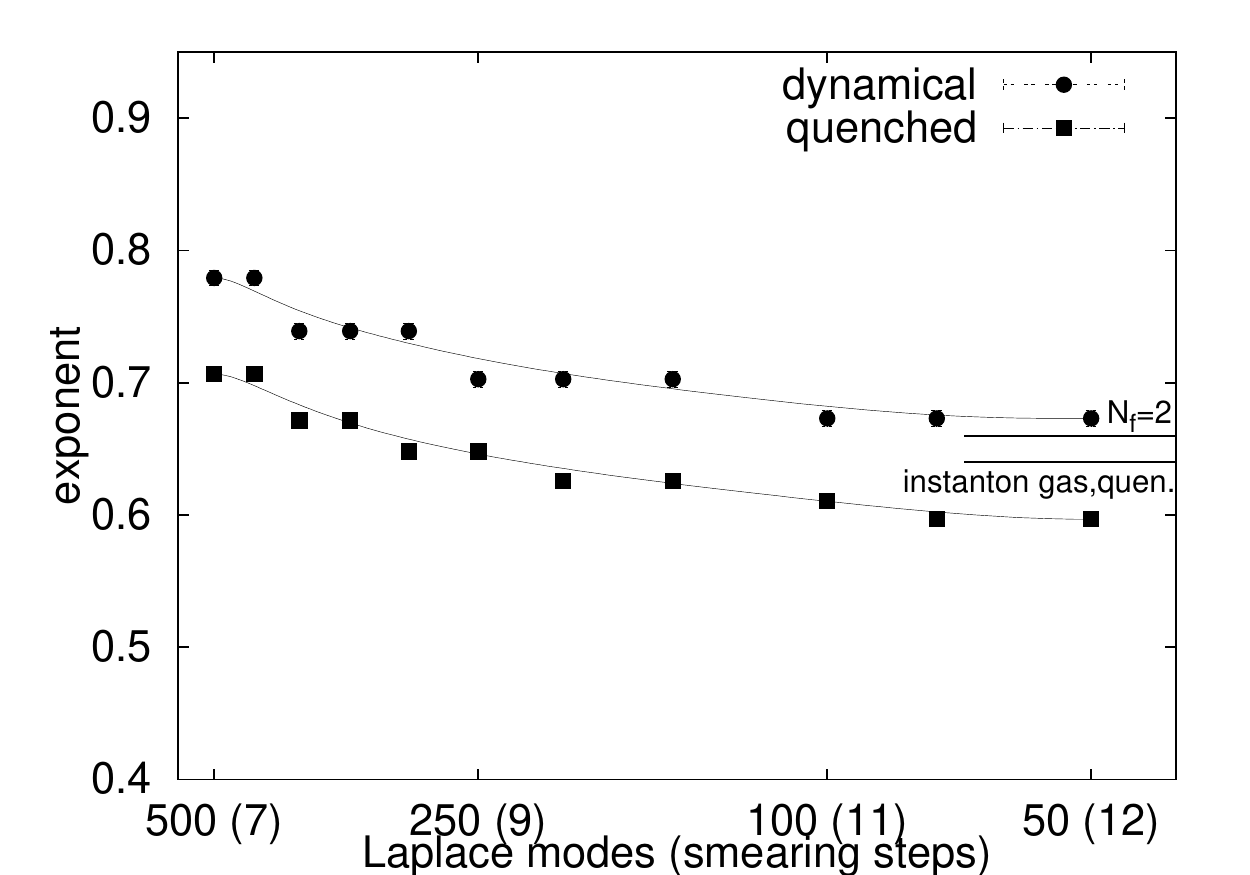}}}
\caption{Exponent $\xi$ for the clusters found by APE smearing (a), Laplace filtering (b) and for a matched filtering of both methods (c). The exponents predicted by the dilute instanton gas have been included in all plots. In the latter plot the measured exponents have a step-like behavior, because the number of smearing steps is an integer quantity and we have to match several numbers of Laplace modes to the same number of smearing steps. The smooth lines are interpolations corresponding to a ``continuous'' smearing parameter. The number of smearing steps (in parentheses) refers to the matching of quenched configurations and is slightly different for the dynamical case. Errors from the ensemble average have been included, but are partly too small to be seen.}\label{fig: matching}
\end{figure}

Of course, we perform this analysis after having matched the filtering parameters as described in the previous section.
The exponent for clusters common to APE  and Stout smearing can be found in Fig.~\ref{fig: APE cluster analysis}.\ Obviously, the exponents of the dynamical configurations lie above the quenched values. The difference of the cluster exponents quenched vs.\ dynamical vanishes for stronger smearing ($\sim$ 30 steps) 
and the exponents settle down to almost the same plateau. So we have reasons to believe that too much smearing destroys the impact of dynamical quarks.

In Fig.~\ref{fig: Laplace cluster analysis} the same analysis is done for Laplace filtering (only). We find a slightly bigger difference of the exponents of dynamical and quenched configurations. Furthermore, this difference remains, unlike for APE smearing, at every stage of filtering. At first sight the value for 50 modes seems to be odd, but this is just an artifact of the incomplete reconstruction of the topological background. For a small number of modes, we get a very spiky structure and we will find many new clusters, if we lower the cut-off. Thus, we get a higher exponent.

To get rid of the ambiguities of APE smearing and Laplace filtering, we perform in Fig.~\ref{fig:Laplace Ape matched cluster analysis} a matched cluster analysis. i.e. we consider only those clusters, which are common to both methods. We use the optimal set of filter parameters according to the maximal values of $\Xi$ (and corresponding to points on the ridge line of Fig.~\ref{fig: Xi}) on the abscissa.

Fig.~\ref{fig:Laplace Ape matched cluster analysis} shows that the \emph{cluster exponent is larger in the dynamical case} than in the quenched case. According to the considerations from above, this 
implies a larger exponent $\beta$ of the size distribution. Hence, very small topological objects become suppressed when quarks are taken into account. 

In the plots we have included the values of the exponent in instanton gases. These have the same ordering with a slightly smaller difference. Generally our measured values of the exponent are not far off the instanton gas values (which was not the case in $SU(2)$ \cite{Bruckmann2007c}).
\begin{figure}
\centering
\includegraphics[trim= 0 0 30 0,width=0.43\textwidth]{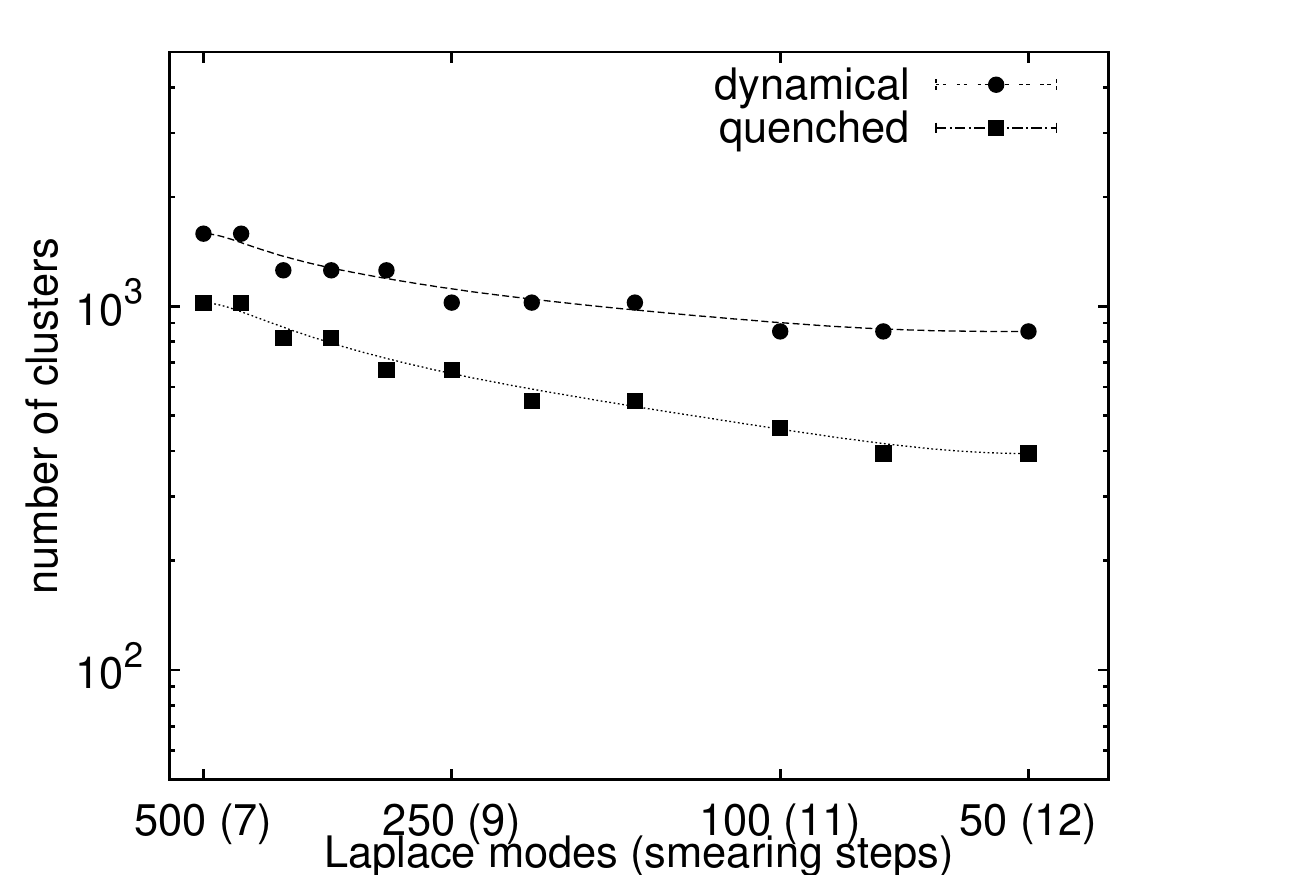}
\caption{Total number of distinct clusters for a constant fraction $f=0.0755$ of points lying above the cut-off with smooth interpolation.}\label{fig:absolute number of matched clusters}\vspace*{-12pt}
\end{figure}

Another interesting quantity is the absolute number of clusters above a certain cut-off, see Fig.~\ref{fig:absolute number of matched clusters}. In the dynamical case we find almost twice as many clusters. Hence, we observe a \emph{more fragmented topological structure in the presence of dynamical quarks}.

Our findings can be compared to the results obtained by the Adelaide group \cite{Moran2008a}. They have observed an ``increasing density of nontrivial field configurations'' and a suppression of small instantons.
Using a completely different approach, we come to the same conclusion. The advantages of our analysis are that our dynamical fermions have nicer chiral properties than the staggered ones. Moreover, we do not have to postulate any shape function of the topological objects. Therefore, our method works for all topological building blocks and with our matched analysis we can obtain results which are almost free from ambiguities of the filtering process.

\section{Conclusions}

We have performed a comparative study on the local topological structure of quenched and dynamical configurations seen by various filtering methods, namely APE smearing, Stout smearing and Laplace filtering.

We found an almost one-to-one correspondence of the topological charge densities resulting from APE  and Stout smearing, if the filtering parameters are chosen properly. This extends findings in lattice perturbation theory \cite{Bernard2000d} and on non-topological observables for few smearing steps \cite{Capitani2006e}. Our results reflect the local similarity of both methods for up to 50 steps and their strongly correlated topological charge densities.

 As a consequence APE  and Stout smearing can be seen as equal for the investigation of topological objects on the lattice. On the other hand, both methods will therefore suffer from almost the same artifacts and other, independent methods are needed to get results free from ambiguities.

To this end Laplace filtering, which is based on a truncated spectral expansion of the Laplacian, has been investigated in addition to smearing. In the case of mild filtering, we found a qualitative agreement of Laplace filtering and APE smearing. This is in accordance with the quenched $SU(2)$ results of Bruckmann {\em et al.} \cite{Bruckmann2007c}. For $SU(3)$ the agreement is slightly better in the quenched case. However, for the strong filtering case we have found a clear disagreement, which is even more pronounced in the dynamical case. This finding flags a warning that any kind of smearing should be applied with great care.

In the central part of this work the topological clusters, revealed by the different methods, have been investigated. To this end we have analyzed the power-law behavior of topological charge clusters seen by the individual methods and common to both methods.

This analysis shows clearly a larger exponent in the presence of dynamical quarks. In order to interpret this observation, we can relate the cluster exponent to the exponent of the size distribution $d(\rho)\sim \rho^\beta$ of topological objects in corresponding models. Our findings give a larger coefficient $\beta$ in the dynamical case, hence, very small topological objects become suppressed. 

Moreover, the total number of clusters for a constant total cluster volume is substantially higher for dynamical configurations. Thus the topological structure of the vacuum is much more fragmented in the presence of fermions. It would be interesting to further investigate the nature of these topological lumps. 

We would like to thank G. Bali, N. Cundy, E.-M. Ilgenfritz, M. M\"uller-Preussker and S. Solbrig for helpful comments and discussions. Furthermore, we want to thank C.B. Lang, M. Limmer and D. Mohler who generated most of the configurations which were used in this Letter and the LRZ which provided much of the computing time used. FB has been supported by DFG BR 2872/4-1 and FG by SFB TR-55.

\bibliographystyle{elsarticle-num}
\bibliography{Filtering}

\end{document}